\documentclass[12pt]{iopart}
\usepackage{graphicx}
\begin{document}

\title{Cooling dynamics of pure and random Ising chains}

\author{Sei Suzuki}

\address{Department of Physics and Mathematics, Aoyama Gakuin
University, Fuchinobe, Sagamihara 229-8558, Japan}
\ead{sei@phys.aoyama.ac.jp}
\begin{abstract}
Dynamics of quenching temperature is studied in pure and 
random Ising chains. Using the Kibble-Zurek argument, 
we obtain for the pure Ising model that 
the density of kinks after quenching decays as $1/\sqrt{\tau}$ 
with the quench rate of temperature $1/\tau$ for large $\tau$.
For the random Ising model, we show that decay rates of 
the density of kinks and the residual energy are
$1/\ln\tau$ and $1/(\ln\tau)^2$ for large $\tau$ respectively. 
Analytic results for the random Ising model
are confirmed by the Monte-Carlo simulation.
Our results reveal a clear difference between classical and quantum 
quenches in the random Ising chain.
\end{abstract}


\section{Introduction}
The change of a parameter across a phase boundary in a macroscopic system
induces the dynamical phase transition of the system.
If the changing speed of the parameter is sufficiently slow, the 
transition propagates over the whole system. However,
as far as the changing speed is finite, symmetry breaking 
does not take place globally but does locally.
It follows that spatial inhomogeneities emerge after the 
time evolution.
Recent progress in experimental techniques enables to demonstrate 
such a dynamics across the phase transition and to observe imperfections
in the ordered state. 
Greiner \textit{et al}. \cite{bib:Greiner} studied the dynamics across the
quantum phase transition between the superfluidity and the Mott
insulator in the optical lattice system. Sadler \textit{et
al}. \cite{bib:Sadler} 
observed formation of defects after the quantum
phase transition from the paramagnetic state to the ferromagnetic
state in the atomic Bose-Einstein condensate.
Weiler \textit{et al}. \cite{bib:Weiler} observed an evidence for
the creation of 
vortices after the thermal phase transition of the Bose-Einstein
condensation in the atomic Bose gas.

The imperfection of the state after the evolution across the phase
transition decays monotonically with decreasing 
the changing speed of the parameter.
The decay rate depends on the choice of the parameter and 
the character of the associated phase transition.
If the temperature is quenched in the classical system,
the system undergoes the classical and thermal phase transition.
If the strength of the quantum fluctuation
is reduced at zero temperature, the quantum phase transition
rules the character of the state.
In the present paper, we study the dynamics near the thermal phase 
transition. Results are contrasted with those obtained for the
quantum dynamics to reveal that there exists a clear difference
between the thermal phase transition and the quantum phase transition
in the decay rate of an imperfection after the time evolution.
Such quench dynamics across the phase transition are 
closely related to the dynamics of simulated annealing \cite{bib:KGV83}
and quantum annealing \cite{bib:FGSSD94,bib:KN98,bib:FGGLLP01}. 
It is an important issue whether 
quantum annealing performs better than simulated annealing or not
\cite{bib:QABook}.

The dynamics across the phase transition is
well understood by the Kibble-Zurek mechanism \cite{bib:K80,bib:Z85}. 
The scenario of 
the Kibble-Zurek mechanism is described as follows.
Suppose that the system is driven from a disordered state to
an ordered state by changing a parameter with a finite speed.
The order parameter of the system in the initial state
is uniformly zero. 
When the parameter comes close to the critical point,
the healing time is so long that the parameter is
changed further before the system attains the static state. 
Hence the system cannot evolves into the perfectly ordered state
after the parameter passes through the critical point.
The system after evolution consists of domains with
different phases of the order parameter. 
The average of the domain size, i.e., the correlation length
grows with decreasing the changing speed of the parameter.
The growing rate is a universal function of the changing speed.

The pure and random Ising chains are the models
that permit us to investigate their dynamical properties analytically.
Both of these models exhibit the quantum phase transition
in the presence of the transverse field. The dynamics
across the quantum phase transition in the pure Ising chain
in the transverse field was first studied by 
Zurek \textit{et al}. on the basis of the Kibble-Zurek argument 
\cite{bib:ZDZ05}.
They showed that the density $\rho$ of kinks between 
ferromagnetic domains behaves as
\begin{equation}
 \rho^{\rm Q}(\tau) \sim 1/\sqrt{\tau}
\label{eq:DOKPure}
\end{equation}
with the quench rate $1/\tau$ of the transverse field. This result 
is confirmed by the analytic solution of the Schr\"{o}dinger equation
by Dziarmaga \cite{bib:D05}.
As for random systems, 
Dziarmaga applied the Kibble-Zurek argument to the quantum
phase transition of the random Ising chain in the transverse
field and obtained 
density of kinks decaying approximately as \cite{bib:D06}
\begin{equation}
 \rho^{\rm Q}(\tau) \sim 1/(\ln\tau)^2 ,
  \label{eq:DOKCaneva}
\end{equation}
for large $\tau$.
Caneva \textit{et al}. also derived the same decay rate \cite{bib:CFS07},
using the Landau-Zener formula 
and the distribution of excitation gaps at the critical point.
Besides the density of kinks, they also estimated the decay rate of the
residual energy. The result is given by
\begin{equation}
E_{\rm res}^{\rm Q}(\tau) \sim 1/(\ln\tau)^{\zeta} ~,~~~
\zeta\approx 3.4 .
\label{eq:EresCaneva}
\end{equation}

The study for the dynamics across the thermal phase transition
in the Ising system is not necessarily sufficient. 
Laguna and Zurek have studied the Langevin 
dynamics of the order-parameter field in one spatial dimension
\cite{bib:LZ97}. 
However the model they studied does not correspond to the true Ising
model in one dimension, because it involves an unphysical phase
transition. Huse and Fisher have made a theory on residual energy
after quenching temperature in classical random systems \cite{bib:HF86}.
They regarded the system with disorder
as a collection of independent two-level systems,
and derived residual energy which decays as
\begin{equation}
E_{\rm res}^{\rm C}(\tau) \sim 1/(\ln\tau)^2 .
 \label{eq:EresHF}
\end{equation}
Despite of Huse-Fisher's general theory on quenching
in classical random systems, one cannot tell anything about comparison 
between dynamics across quantum phase transition and 
thermal phase transition.
Comparison of eqs. (\ref{eq:EresCaneva}) and
(\ref{eq:EresHF}) is obscure because of a lack of analytical support
on eq. (\ref{eq:EresCaneva}). Density of kinks tells nothing 
since it is not available in the classical dynamics.

The results we obtain are summarized as follows. 
The density of kinks after quenching temperature in 
the pure Ising chain is given by
\begin{equation}
 \rho^{\rm C} \sim 1/\sqrt{\tau}
\end{equation}
for large $\tau$. In the random Ising chain, it behaves as
\begin{equation}
 \rho^{\rm C}(\tau) \sim 1/\ln \tau 
  \label{eq:DOKSSRandom}
\end{equation}
for large $\tau$.
As for residual energy, eq. (\ref{eq:EresHF})
is reproduced for the random Ising chain. 
We emphasize that its derivation uses a manner 
different from the Huse-Fisher's theory.
Our results reveal a clear difference in the decay rate of
the density of kinks between the quantum quench and the classical quench.

This paper is organized as follows. At first, we briefly review 
the Kibble-Zurek argument in the next section. After that,
we study quench dynamics of the
pure Ising chain and derive the decay rate of the density of kinks
in sec. \ref{sec:Pure}. 
We then reveal logarithmic decay rates 
of the density of kinks and the residual energy for the random Ising
chain in sec. \ref{sec:Random}.
We also show results of Monte-Carlo simulation for the random Ising
chain there. The paper is concluded in sec. \ref{sec:Conclusion}.

\section{Kibble-Zurek argument}\label{sec:KZ}
Let us consider a ferromagnetic system with the 
critical temperature $T_c$. 
In the Kibble-Zurek argument, the correlation length and the
relaxation time of the system with a fixed temperature are
quantities of importance. Both quantities are functions of
temperature and increase with decreasing temperature toward $T_c$. 
We denote the correlation length and the relaxation time 
by $\xi(T)$ and $\tau_{\rm r}(T)$ respectively.

Now we consider quenching temperature with time $t$ as
$T(t) = T_c \left(1-t/\tau\right)$
where time is assumed to evolve from $-\infty$ to $\tau$ and
$1/\tau$ stands for the quench rate. 
We assume that the system is in its equilibrium state
initially. When the temperature is sufficiently high, the
system almost maintains its equilibrium since the relaxation
time is short. However, when the temperature is close to 
$T_c$, the temperature decreases further before the system
attains the equilibrium. Thus the system cannot possess the
complete ferromagnetic order and contains domain walls when
the temperature goes below $T_c$. Once the domain structure
forms, it should preserve until the temperature reaches absolute
zero.
The size of the domain is represented by the 
correlation length $\hat{\xi}$ of the state when the temperature
passes $T_c$. An argument described below provides an estimation
of $\hat{\xi}$ for a given $\tau$.

We introduce an equality:
\begin{equation}
 \tau_{\rm r}(T(\hat{t})) = |\hat{t}| .
\label{eq:KZEquality}
\end{equation}
This equality defines the time $\hat{t}$ at which the 
relaxation time is equal to the remaining time to the
critical temperature. At a later time until $t=\tau$, 
the system cannot attain the equilibrium since the relaxation time is
longer than the remaining time. 
Suppose here that the system stays in 
the equilibrium at $\hat{T}\equiv T(\hat{t})$ and
does not evolve any more after $t$ passes $\hat{t}$. 
Then the correlation length of the
state at $t=\tau$ is approximated by $\hat{\xi}\approx\xi(\hat{T})$.
Since one can express $T$ in terms of $\xi$ from the
expression of $\xi(T)$, the left hand side of eq. (\ref{eq:KZEquality})
is written in terms of $\hat{\xi}$. 
The right hand side, on the other hand, 
is written as $\tau|\hat{T}-T_c|/T_c$, which is also expressed in
terms of $\hat{\xi}$. Thus we obtain the equation of
$\hat{\xi}$ from eq. (\ref{eq:KZEquality}).
The solution of this equation yields $\hat{\xi}$ as a function of $\tau$.

\section{Pure Ising chain}\label{sec:Pure}
We consider the simple pure Ising model in one dimension:
$H = - \sum_i\sigma_i\sigma_{i+1}$.
Although this model does not exhibit the 
phase transition at any finite temperature, the
ground state possesses the complete ferromagnetic order. 
Hence one can regard the critical temperature as $T_c=0$.
Denoting the inverse of temperature by $\beta$,
an expression of the correlation length is given by
\begin{equation}
 \xi(T) = 1/\ln\coth\beta \approx \frac{1}{2}e^{2\beta} ,
  \label{eq:xi-pure}
\end{equation}
where the lattice constant is assumed to be the unit of length.
We note that the right hand side is the expression valid at low
temperature, i.e., $T\ll 1$. In order to discuss the dynamics of the
present system, we
assume the Glauber model\cite{bib:G63}. Then, the relaxation
time for a fixed temperature is given by 
\begin{equation}
 \tau_{\rm r}(T) = 1/(1 - \tanh 2\beta) 
  \approx \frac{1}{2}e^{4\beta} \approx
  \frac{1}{2}\xi(T)^2 ,
\label{eq:tau-pure}
\end{equation}
where the approximation signs are valid at low temperature.
Thus the correlation length and the relaxation time 
grow with decreasing temperature toward $T_c = 0$.

From now we discuss the dynamics of quenching temperature
according to the Kibble-Zurek argument. 
We assume the quench schedule:
\begin{equation}
 T(t) = - t/\tau 
  \label{eq:Schedule1}
\end{equation}
instead of the one in the previous section because of $T_c = 0$. 
We also assume
that the time $t$ evolves from $-\infty$ to $0$ and the inversed quench
rate $\tau$ is large, i.e., $\tau\gg 1$. Equation (\ref{eq:KZEquality})
defines the approximate time $\hat{t}$ at which the evolution of 
the system stops.
Using eqs. (\ref{eq:xi-pure}) and (\ref{eq:Schedule1}),
the time $\hat{t}$ is written as $\hat{t} \approx -2\tau/\ln 2\hat{\xi}$,
where $\hat{\xi}$ is the correlation length at $\hat{T}=T(\hat{t})$.
We remark that
the low temperature expression of $\xi$ is allowed
as far as $\tau$ is large enough because $\hat{T}$ is small.
Equations (\ref{eq:KZEquality}) and (\ref{eq:tau-pure}) yield
an equation of $\hat{\xi}$ as 
$\hat{\xi}^2 \approx 4\tau/|\ln 2\hat{\xi}|$.
This equation cannot be solved analytically. However
$\ln\hat{\xi}$ is a gentle function of $\hat{\xi}$
compared to $\hat{\xi}^2$. Hence $\hat{\xi}$ is almost 
proportional to $\sqrt{\tau}$.
The inverse of correlation length corresponds to 
the density of kinks approximately. 
It follows that the density of kinks in the final state is
estimated as 
\begin{equation}
 \rho\approx \frac{1}{\hat{\xi}} = 
\frac{(|\ln 2\hat{\xi}|)^{1/2}}{2\sqrt{\tau}} .
\end{equation}
Thus one finds that the density of kinks of the final state 
is proportional to $1/\sqrt{\tau}$ 
as far as the logarithmic correction is ignored.
The logarithmic correction is not essential indeed.
To verify this, we consider a modification of the Kibble-Zurek
argument as follows. We may employ another
equality, $\tau_{\rm r}(T(\tilde{t})) = \tau$, instead of
eq. (\ref{eq:KZEquality}). 
This equality defines the
time $\tilde{t}$ at which the relaxation time exceeds
the inverse of quench rate. we can consider that 
the evolution of the system almost stops at $\tilde{t}$ and
construct the argument same as that in the previous section.
By this argument, we obtain $\rho\sim 1/\sqrt{\tau}$ without
the logarithmic correction.

\section{Random Ising chain}\label{sec:Random}
The random Ising chain is represented by
\begin{equation}
 \mathcal{H} = - \sum_i J_i \sigma_i \sigma_{i+1} .
\label{eq:H}
\end{equation}
In our study, the coupling constant $\{J_i\}$ is 
drawn randomly from the uniform distribution 
between $0$ and $1$, namely $P(J_i)=1$ for $J_i\in [0,1]$ and
$P(J_i)=0$ otherwise.
This model corresponds to the one studied in 
refs.\cite{bib:D06,bib:CFS07}. 

The correlation function between sites $i$ and $i+k$ 
with fixed $\{J_i\}$ in the equilibrium at a fixed temperature 
is given by $\langle \sigma_i\sigma_{i+k}\rangle =
\prod_{j=i}^{i+k-1}\tanh\beta J_j$. 
Taking the average over randomness, the correlation function 
in the thermodynamic limit is obtained as
$[\langle \sigma_i\sigma_{i+k}\rangle]_{\rm av} 
 = (\ln\cosh\beta/\beta)^k$.
From this formula of the correlation function,
one can obtain an explicit
expression of the correlation length: 
\begin{equation}
 \xi (T) = [\ln (\beta/\ln\cosh\beta) ]^{-1}
\approx  \beta/\ln 2 .
\label{eq:Xi}
\end{equation}
Note that the right hand side is the low temperature expression.

The energy of the system with fixed $\{J_i\}$ is 
written as 
$\langle \mathcal{H}\rangle = - \sum_i J_i \tanh\beta J_i$.
The average over randomness yields an expression
of the energy per spin at low temperature in the thermodynamic limit:
\begin{equation}
 \varepsilon = \lim_{N\to\infty}
\frac{[\langle\mathcal{H}\rangle]_{\rm av}}{N} \approx 
- \frac{1}{2} + 
  \frac{1}{\beta^2}\frac{\pi^2}{24} .
\label{eq:EperSpin}
\end{equation}
We remark that the ground state energy is $-\frac{1}{2}$.

The relaxation time is available in ref.\cite{bib:DB80}
by Dhar and Barma. It is given by
$\tau_{\rm R} = 1/(1 - \tanh 2\beta)$. 
The low temperature expression is
\begin{equation}
 \tau_{\rm r}(T) \approx \frac{1}{2}e^{4\beta} 
\approx \frac{1}{2}e^{(4\ln 2)\xi (T)} .
\label{eq:TauR}
\end{equation}
As is the case with the pure Ising chain, 
the critical temperature of the present model is $T_c = 0$.

Let us consider that the temperature is lowered 
according to the schedule given by eq. (\ref{eq:Schedule1}).
We impose eq. (\ref{eq:KZEquality}) to define the time $\hat{t}$
at which the evolution keeping equilibrium breaks.
From eq. (\ref{eq:Xi}), the time relates with the correlation 
length by $|t|=\tau/(\xi\ln 2)$. Applying this relation and 
eq. (\ref{eq:TauR}) to eq. (\ref{eq:KZEquality}), 
we obtain an equation of $\hat{\xi}$:
\begin{equation}
 \hat{\xi}=\frac{1}{4\ln 2}\left(\ln\tau - 
\ln\frac{\hat{\xi}\ln 2}{2}\right) .
\label{eq:XiEquation}
\end{equation}
This equation cannot be solved analytically.
However, since $(\ln\hat{\xi})/\hat{\xi} \to 0$ for $\hat{\xi}\to 0$,
we find that $\hat{\xi}$ is almost proportional to $\ln\tau$
when $\tau\gg 1$.
Equation (\ref{eq:XiEquation}) leads to an estimation of 
the density of kinks, 
\begin{equation}
 \rho \approx \frac{1}{\hat{\xi}} \approx \frac{4\ln 2}{\ln\tau - 
  \ln \frac{\hat{\xi}\ln 2}{2}} .
\label{eq:DOKLin}
\end{equation}
The second term in the denominator is negligible for a
sufficiently long $\tau$ as mentioned above.
Hence eq. (\ref{eq:DOKSSRandom}) is derived. 

The residual energy is also estimated from the energy 
at $T=\hat{T}$. Using eqs. (\ref{eq:TauR}) and 
(\ref{eq:Schedule1}), eq. (\ref{eq:KZEquality}) is 
rewritten as
\begin{equation}
\frac{1}{2}e^{4\hat{\beta}}=\tau/\hat{\beta} ,
\label{eq:BetaEquation}
\end{equation}
where we defined $\hat{\beta} \equiv 1/\hat{T}$.
This equation is followed by
$\hat{\beta} = \frac{1}{4}\ln \tau - 
\frac{1}{4}\ln(\hat{\beta}/2)$.
Substituting this for $\beta$ in eq. (\ref{eq:EperSpin}),
we obtain the residual energy per spin as
\begin{equation}
 \varepsilon_{\rm res} = \frac{2\pi^2}{3}\frac{1}{(\ln \tau - 
\ln(\hat{\beta}/2))^2} .
\label{eq:EresLin}
\end{equation}
Since the second term in the denominator 
is negligible for large $\tau$, hence we obtain 
eq. (\ref{eq:EresHF}).

We next consider a logarithmic schedule:
\begin{equation}
 T (t) = \frac{T_0}{1 + a\ln(-\frac{T_0\tau}{t})} ,
\label{eq:LogSchedule}
\end{equation}
where $T_0$ and $a$ are positive numbers.
In this schedule, the temperature is reduced from 
$T_0$ at $t=-T_0\tau$ to $0$ at $t=0$.
Using eq. (\ref{eq:LogSchedule}) with eqs. (\ref{eq:KZEquality}) and
(\ref{eq:TauR}), one obtains the equation of $\hat{\beta}$ as
$\frac{1}{2}e^{4\hat{\beta}} = T_0\tau
\exp\{1/a - (T_0/a)\hat{\beta}\}$.
This equation can be solved analytically and yields 
$\hat{\beta} = \ln(2e^{1/a}T_0\tau)/(4+T_0/a)$.
From eq. (\ref{eq:Xi}), one obtains the expression
for density of kinks as
\begin{equation}
 \rho \approx \frac{1}{\hat{\xi}} \approx
  \frac{(4 + \frac{T_0}{a})\ln 2}
  {\ln\tau + \ln (2T_0) + \frac{1}{a}} .
\label{eq:DOKLog}
\end{equation}
This expression is reduced to eq. (\ref{eq:DOKSSRandom}) for 
$\tau \to\infty$. The expression of the residual energy per spin
is obtained as
\begin{equation}
 \varepsilon_{\rm res} \approx \frac{\pi^2}{24}
  \frac{(4 + \frac{T_0}{a})^2}
  {\left(\ln\tau + \ln (2T_0) + \frac{1}{a}\right)^2} ,
\label{eq:EresLog}
\end{equation}
which yields eq. (\ref{eq:EresHF}) for $\tau\to\infty$.
These results imply that 
asymptotic behaviors of the density of kinks and the
residual energy for $\tau\to\infty$ are insensitive 
to the schedule of quenching temperature.

If the distribution of $J_{ij}$ has a finite positive lower-bound,
decay rates of the density of kinks and the residual energy are
the same and obey the power law. To show this, we 
suppose $J_{ij}\in [J_0, 1]$ with $0 < J_0 < 1$. 
Then the correlation length is written as
$\xi\approx \beta\mathit{\Delta}J e^{2\beta J_{0}}$, where we defined 
$\mathit{\Delta}J=1-J_0$. The energy per spin is given by
$\varepsilon\approx\varepsilon_g + 
\frac{1}{\beta\mathit{\Delta}J}J_0e^{-2\beta J_0}$, where
$\varepsilon_g = -\frac{1}{2\mathit{\Delta}J}(1-J_0^2)$ is the ground-state
energy. The relaxation time does not change by
introducing $J_0$ \cite{bib:DB80}. It follows that the condition in the 
Kibble-Zurek argument, eq. (\ref{eq:KZEquality}), yields the 
same equation of $\beta$ as eq. (\ref{eq:BetaEquation}).
Using its solution for large $\tau$,
We obtain $\varepsilon_{\rm res}\approx J_0\rho$ and
$\rho\sim \tau^{-J_0/2}/\ln\tau$. 
As for the quantum quench, the introduction of a finite positive
$J_0$ does not change the universality of quantum phase transition
\cite{bib:F95}. It follows that the decay rate of the density of kinks
is logarithmic and given by eq. (\ref{eq:DOKCaneva}).
Hence the classical quench reduces the density of kinks
faster than the quantum quench in this case.

We confirm results of the random Ising chain
by the Monte-Carlo simulation for systems with $500$ spins.
The temperature is lowered according to the 
linear schedule. 
We choose the initial condition for the temperature as
$T=5$ at $t = -5\tau$. The coupling constant $J_{ij}$ is
drawn from $[0,1]$ uniformly.
In order to take an average with respect to 
randomness of the system, we generated 100 configurations of coupling
constants $\{J_{i}\}$. For each 
configuration, we perform quenching temperature 500 times.

\begin{figure}[tbp]
\begin{center}
 \includegraphics[width=8cm,clip]{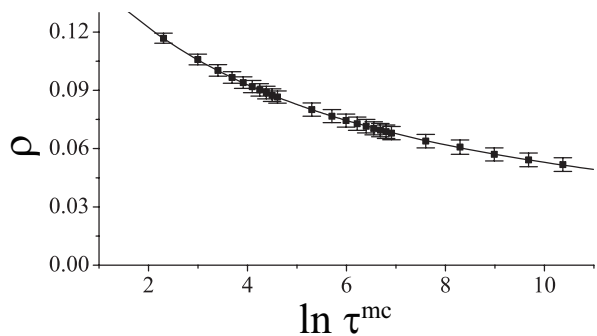}
\end{center}
 \caption{The density of kinks after cooling with the linear
schedule. Square symbols are obtained by the Monte-Carlo simulation.
The fitting curve is given by
$\rho = q/\hat{\xi}$ with the correlation length $\hat{\xi}$ determined by
eq. (\ref{eq:XiEquation}) with $\tau=p\tau^{\rm mc}$.
Parameters $p$ and $q$ are determined by denotes of the method
of least squares. 
}
\label{fig:DOK}
\end{figure}
\begin{figure}[tbp]
\begin{center}
 \includegraphics[width=8cm,clip]{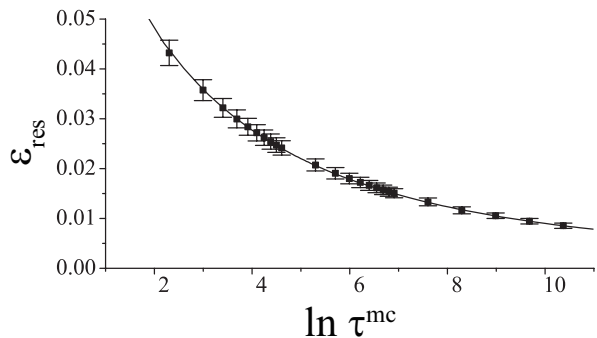}
\end{center}
 \caption{The residual energy obtained by Monte-Carlo simulation 
(square symbols). The fitting function is given by 
$\varepsilon_{\rm res} = \frac{\pi^2}{24}
/(r\hat{\beta})^2$
with $\hat{\beta}$, the solution of eq. (\ref{eq:BetaEquation})
with $\tau = p\tau^{\rm mc}$.
The fitting parameter $r$ is determined by the method
of least squares with $p$ given from fitting of density of kinks.
}
\label{fig:Eres}
\end{figure}

Square symbols in Fig. \ref{fig:DOK} and Fig. \ref{fig:Eres}
show the density of kinks and the residual energy per spin respectively 
obtained by the Monte-Carlo simulation. The density of kinks
is defined by $
\rho = [\frac{1}{2N}\sum_i(1 - \langle \sigma_i\sigma_{i+1}
\rangle)]_{\rm av}$,
where
$\langle\cdots \rangle$ denotes the expectation value
with respect to the state after simulated annealing and
$[\cdots]_{\rm av}$ means the average over configurations of
coupling constants. 

In order to obtain fitting curves for the density of kinks
and the residual energy per spin, 
we need to modify eqs. (\ref{eq:DOKLin}) and (\ref{eq:EresLin}). 
First, we have to care about the difference 
in the unit of time between the Glauber's dynamics and
the Monte-Carlo dynamics. Then we bring up the relation, 
$\tau = p\tau^{\rm mc}$, between the inverse of 
quench rate $\tau$ in the Glauber's dynamics and 
$\tau^{\rm mc}$ in the Monte-Carlo dynamics, where $p$ is
an adjustable parameter.
Next, we relate the density of kinks $\rho$ to 
the correlation length $\xi$ by $\rho = q/\hat{\xi}$,
where $\hat{\xi}$ is the solution of eq. (\ref{eq:XiEquation}).
The parameter $q$ tunes the inverse of correlation length to
the density of kinks.
Finally, we propose an ansatz that the residual energy is represented
by $\varepsilon_{\rm res} = (\pi^2/24)
/(r\hat{\beta})^2$ with an adjustable parameter $r$, 
where $\hat{\beta}$ is given from 
eq. (\ref{eq:BetaEquation}) with $\tau=p\tau^{\rm mc}$.

Parameters $p$ and $q$ are
determined by the method of least squares described as follows. 
The Monte-Carlo simulation yields a set of data 
$(\rho_i,\tau^{\rm mc}_i)$, where $\rho_i$ is the
mean value produced by Monte-Carlo simulations.
 The fitting function given by
eq. (\ref{eq:DOKLin}) but $\rho = q/\hat{\xi}$ and
$\tau=p\tau^{\rm mc}$ does not yield $\rho$ for given $\tau^{\rm mc}$
analytically. Then we regard $\tau^{\rm mc}$ as a function
of $\rho$ and define the error between Monte-Carlo
data and the fitting function by 
$S(p,q) \equiv \sum_{i} (\ln \tau^{\rm mc}(\rho_i) - \ln\tau^{\rm mc}_i)^2
/\sigma_i^2$
where $\tau^{\rm mc}(\rho_i)$ is the value of
the fitting function and $\tau^{\rm mc}_i$ is given from the
Monte-Carlo data for $\rho_i$. $\sigma_i^2$ is the dispersion
of $\rho_i$. We assume that the relative values of the dispersion
between different $i$'s are the same in $\rho$ and $\ln\tau^{\rm mc}$.
By minimizing $S(p,q)$, we fix the value of $p$ and $q$.
The errors of $p$ and $q$ are given by 
$\sigma_p^2 = \frac{2 S}{m-2}\frac{\partial^2 S}{\partial q^2}
\frac{1}{D}$ and
$\sigma_q^2 = \frac{2 S}{m-2}\frac{\partial^2 S}{\partial p^2}
\frac{1}{D}$, where  $m$ is the number of Monte-Carlo data and
$D=\frac{\partial^2 S}{\partial p^2}\frac{\partial^2 S}{\partial q^2}-
(\frac{\partial^2 S}{\partial p\partial q})^2$.
The other parameter $r$ is determined by the method of
least squares with the error $S'(r) \equiv
\sum_i(\varepsilon(\tau^{\rm mc}_i)
-\varepsilon_i)^2/\sigma_{\varepsilon i}^2$ between 
the value of fitting function, eq. (\ref{eq:EresLin}),
with $\hat{\beta}_i$ given by eq. (\ref{eq:BetaEquation})
with $\tau_i=p\tau^{\rm mc}_i$ and the value of
Monte-Carlo data $\varepsilon_i$. $\sigma_{\varepsilon i}^2$ is the dispersion 
of $\varepsilon_i$. $p$ is fixed by fitting of $(\rho,\tau^{\rm mc})$.
The error of $r$ is given by 
$\sigma_r^2 = \frac{S'}{m-1}(\frac{\partial^2 S'}{\partial r^2})^{-1}$.
The obtained values of $p$, $q$ and $r$ are
$q\approx 0.241\pm 0.001$, $p\approx 22.0\pm 0.9$, and 
$r \approx 2.14\pm 0.16$.

Figure \ref{fig:DOK} and \ref{fig:Eres} shows that results of 
Monte-Carlo simulation on the density of kinks and the residual energy
are excellently fitted by the curves 
made from eq. (\ref{eq:DOKLin}) and (\ref{eq:EresLin}) respectively. 
Therefore the analytic results on the basis of the Kibble-Zurek argument
is confirmed by the Monte-Carlo simulation.

\section{Conclusion}\label{sec:Conclusion}
We studied the
dynamics of quenching temperature of pure and random 
Ising chains, on the basis of the Kibble-Zurek argument. 
We showed for the pure Ising chain that 
the density of kinks after quenching
decays as $1/\sqrt{\tau}$ for large $\tau$. 
As for the random Ising chain with $J_i\in [0,1]$, 
the density of kinks and the residual energy
decay as $1/\ln\tau$ and $1/(\ln\tau)^2$ for large $\tau$ respectively. 
Results for the random Ising chain were confirmed by
the Monte-Carlo simulation.
Comparing our results on the density of kinks with known results 
for the quantum quench, densities of kinks after the classical quench
and the quantum quench decay with the same power of $\tau$
in the pure system. 
As for the random Ising chain, 
the power of $\ln\tau$ by the 
quantum quench is twice as large as that by the classical quench.
The difference between the quantum quench and the classical quench
is substantial.

The classical quench and the quantum quench toward the ground state 
correspond to simulated annealing and quantum annealing
respectively of an optimization problem. 
The random Ising chain studied in the present paper
provides an optimization problem with the trivial solution.
However it is a non-trivial problem for simulated annealing
and quantum annealing since their dynamics toward the
solution respond to randomness and exhibit the slow relaxation. 
Our results reveal that the random Ising chain is a solid
example for which quantum annealing certainly performs better than
simulated annealing.


The author thanks T. Caneva, G. E. Santoro, and H. Nishimori for
fruitful discussions. The present work was partially supported
by CREST, JST.

\end{document}